\documentstyle[prl,aps,epsfig]{revtex}
\newcommand{\ber}{\begin{eqnarray}}
\newcommand{\eer}{\end{eqnarray}}
\newcommand{\be}{\begin{eqnarray}}
\newcommand{\ee}{\end{eqnarray}}

\begin{document}
 \hbadness=10000
\title{Chemical equilibration  of U(1)  charged particles in a thermal environment }
\author{ Krzysztof Redlich$^{a,b}$, Volker Koch${^{c,d}}$, and  Ahmed Tounsi$^e$}

\address{
$^{\rm a}$ Theory Division, CERN, CH-1211 Geneva 23,
Switzerland\\
$^{\rm b}$ Institute of Theoretical Physics, University of Wroc\l
aw, PL-50204 Wroc\l aw, Poland \\
$^{\rm c}$  Nuclear Science Division, Lawrence Berkeley National
Laboratory, 1 Cyclotron Road, Berkeley, CA 94720, USA\\
$^d$ Gesellschaft f\"ur Schwerionenforschung, GSI, D-64291
Darmstadt,
Germany \\
 $^{\rm e}$Laboratoire de Physique Th\'eorique et Hautes
Energies,
Universit\'e Paris 7, 2 pl. Jussieu, F--75251 Cedex 05, France\\
}

%
\maketitle
\begin{abstract}
We discuss chemical equilibration of particles carrying
non-vanishing quantum numbers related with U(1) internal symmetry.
We construct the transport equation for the time evolution of
particle multiplicities and their probability functions. The
solution of these equations is obtained in different limiting
cases. It is argued that a U(1) charged particles, dependent on
thermal conditions inside a fireball,  approaches  different
equilibrium limits. The  differences between kinetics of
abundantly and rarely produced particles are explained.
\end{abstract}

\section{introduction}
One of the crucial  questions being addressed   in the context of
heavy ion collisions is the equilibration of the high density QCD
medium created during these collisions
\cite{satz,sh,stachel,stock}. This question has been considered on
different levels: by analysing physical conditions required for a
perturbative  QCD medium  to reach equilibrium \cite{rolf} or by
studying the level of particle equilibration in the final state
\cite{CLK,braun1}. From the theoretical point of view to discuss
equilibration one needs to formulate the kinetic equation for
particle production and evolution. In the partonic medium this
requires, in general, the formulation of a transport equation
involving colour degrees of freedom and a non-abelian structure of
QCD dynamics \cite{elze86,blaizot1,qun}. In the hadronic medium,
on the other hand, one needs to account for the charge
conservations related with U(1) internal symmetry
\cite{our1,lin,our2}.

In this article we will discuss  the formulation of kinetic
equations for U(1) charged particles  produced in a thermal
environment. We will indicate the importance of the  conservation
laws in the time evolution and chemical equilibration of particle
multiplicities and their probability distributions. In particular
we will argue that the constraints imposed by U(1) charge
conservation are of crucial importance for rarely produced charged
particle species.

\section{Kinetic master equation for probabilities with U(1) symmetry constraints}

To study equilibration in a hadronic medium one  introduces a
kinetic model that takes into account the production and
annihilation of particle--antiparticle pairs $c\bar c$   carrying
U(1) quantum numbers. It is also assumed that      particles $c$
are produced according to a binary process $ab \to c\bar c$ and
that all particle  momentum distributions are thermal  and
described by Boltzmann statistics. The U(1) charge neutral
particles $a$ and $b$  are constituents of a thermal fireball  of
temperature $T$ and volume $V$. We will consider the time
evolution and equilibration of particles $c $  inside a  thermal
fireball, taking into account the constraints imposed by U(1)
symmetry. First we formulate a general master equation for the
probability distribution of particle multiplicity in the medium
with vanishing net charge and consider its properties and
solutions. We will then discuss two limiting cases of abundant and
rare particle production.  Finally the rate equation will be
extended to the situation where there is a net U(1) charge in the
thermal fireball.

\subsection{Rate equation for particle  multiplicity distribution}

Consider $P_{N_c}(t)$ as the probability to find $N_c$ particles
$c$, where $0\leq N_c\le \infty$. This probability will obviously
change in time owing to  production
 $a b\to c\bar c$
and absorption $c\bar c \to ab$ processes. The probability
$P_{N_c}$ tends to increase in time, following  the transition
from $N_c-1$ and $N_c+1$ states to  the $N_c$ state. It also tends
to decrease since the state $N_c$ makes transitions to $N_c+1$ and
$N_c-1$ (see Fig.~1).
 The transition
probability per unit time from  $N_c+1\to N_c$  is given by the
product of the probability $L/V$ that the single reaction $c\bar c
\to ab$ takes place multiplied  by the number of possible
reactions which is, $(N_c+1)(N_{\bar c}+1)$. In the case when the
U(1) charge carried by particles $c$ and $\bar c$ is exactly and
locally conserved, that is if $N_c+N_{\bar c}=0$; then this number
is just $(N_c+1)^2$. Similarly the transition probability from
$N_c\to N_c+1$ is described by $G\langle N_{a}\rangle\langle
N_{b}\rangle /V$, where one assumes that
 particles $a$ and $b$ are not correlated and their multiplicity
 is governed by the thermal averages. One also assumes that
 the multiplicity of $a$ and $b$ is not affected by the $ab \to c\bar
 c$ process.

\begin{figure}[htb]
 {
\includegraphics[width=39.9pc, height=10.pc]{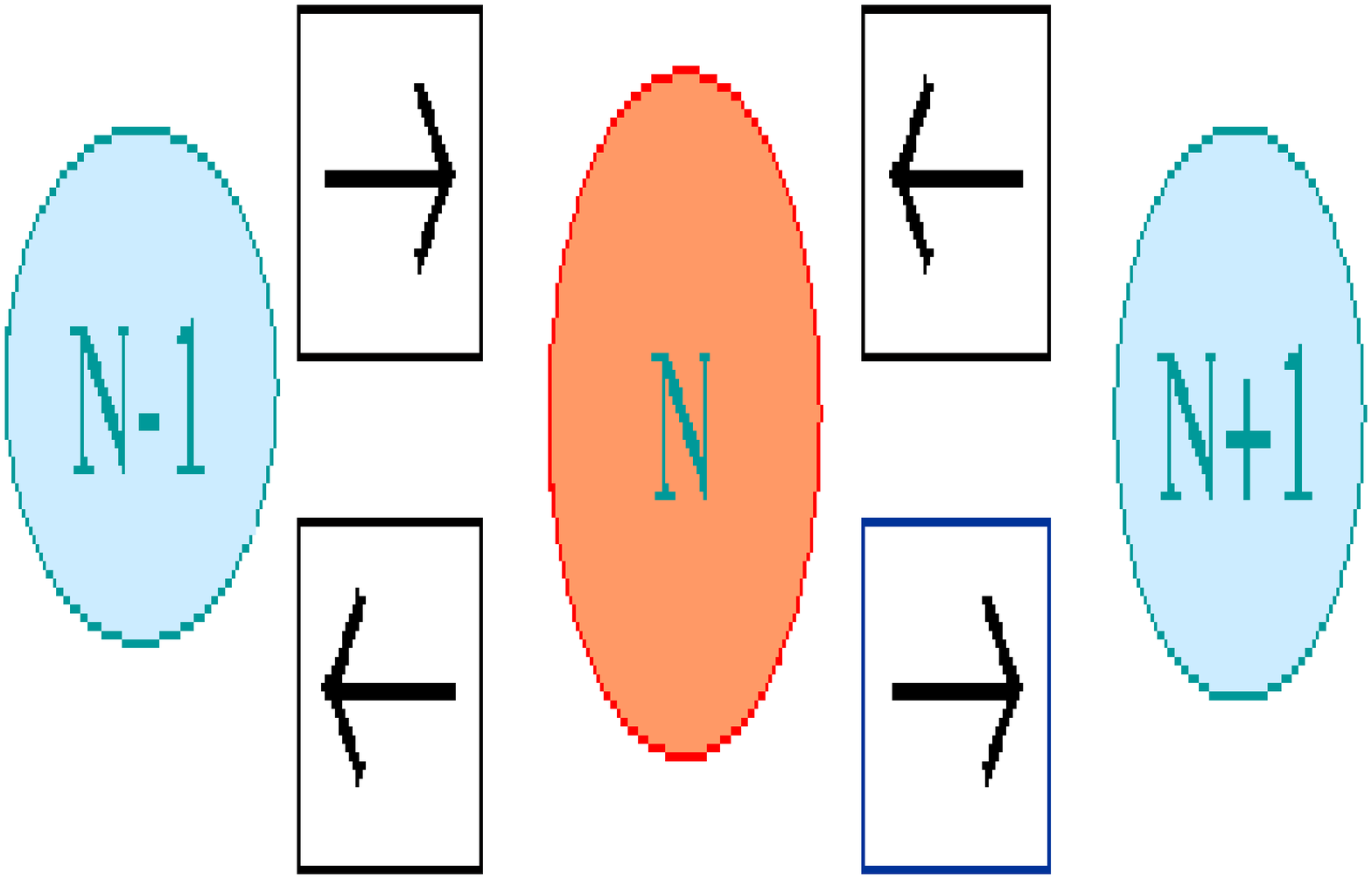}}\\
\vskip -0.8cm { \caption{Schematic view of the master equation for
the probability $P_N(\tau )$ due to  $a b\to c\bar c$ and the
inverse process. }}
\end{figure}
 \noindent The master equation for the time evolution
of the probability $P_{N_c}(\tau )$ can be written in the
following form \cite{our1}:
\ber \frac{dP_{N_c}}{d\tau}&=&{G\over V} \langle
N_{a}\rangle\langle N_{b}\rangle P_{N_c-1} +\frac {L}{V} (N_c+1)^2
P_{N_c+1}
 \nonumber \\
&-& {G\over V}\langle N_{a}\rangle\langle  N_{b}\rangle
P_{N_c}~~~- \frac {L}{V} N_c^2 P_{N_c} . \label{eq1} \eer
The first two terms in Eq.~(\ref{eq1}) describe the increase of
$P_{N_c}(\tau )$
 due to the transition from $N_c-1$ and $N_c+1$ to the $N_c$ state.
The last two terms, on the other hand, represent the decrease of
the probability function due to the transition from $N_c$ to
$N_c+1$ and $N_c-1$ states respectively.

Multiplying the above equation by $N_c$ and summing over $N_c$,
one obtains the general kinetic equation for the time evolution of
the  average number $\langle N_c\rangle\equiv \sum_{N_c=0}^\infty
N_cP_{N_c}(\tau )$ of particles $c$ in a system. This equation
reads:

\begin{eqnarray}
\frac{d\langle N_c\rangle }{d\tau}={G\over V} \langle
N_{a}\rangle\langle N_{b}\rangle - \frac{L}{V} \langle N_c^2
\rangle . \label{normal2}
\end{eqnarray}

The above equation cannot be obviously   solved  analytically as
it connects particle multiplicity $\langle N_c\rangle$ with its
second moments $\langle N_c^2\rangle$. However, the solution can
be obtained in two limiting situations: {\bf i)} for abundant
production of $c$ particles, that is when $\langle N_c\rangle\gg1$
or {\bf ii)} in the opposite limit of rare particle production
corresponding to $\langle N_c\rangle\ll 1$. Indeed, since
\begin{eqnarray}
{\langle N_c^2\rangle }=\langle N_c \rangle ^2+\langle \delta
N_c^2 \rangle, \label{5}
\end{eqnarray}
where $\langle \delta N_c^2 \rangle$ represents the  fluctuations
of the number of particles  $c$, one can make the following
approximations:

\noindent {\bf i)} for  $\langle N_c\rangle \gg1$ one has,
$\langle N_c^2 \rangle \approx \langle N_c\rangle ^2, $ and
Eq.~(2) obviously reduces to the well known form \cite{rafelski}:
\begin{eqnarray}
\frac{d\langle N_c\rangle }{d\tau}\approx {G\over V} \langle
N_{a}\rangle \langle N_{b}\rangle - \frac{L}{V} \langle N_c
\rangle ^2; \label{normal1}
\end{eqnarray}

\noindent however, for rare  production,  particles $c$ and $\bar
c$ are strongly correlated; then,

\noindent {\bf ii)}   $\langle N_c\rangle\ll 1$ and  $\langle
N_c^2 \rangle \approx \langle N_c\rangle$; consequently  Eq.~(2)
takes the form:
\begin{eqnarray}
\frac{d\langle N_c\rangle }{d\tau}\approx {G\over V} \langle
N_{a}\rangle \langle N_{b}\rangle - \frac{L}{V} \langle N_c
\rangle, \label{normal3}
\end{eqnarray}
where the absorption term depends only linearly, instead of
quadratically, on the particle multiplicity.

From the above discussion it is thus clear that depending on the
thermal conditions of the system (that is its  volume and the
value of the temperature)
%
we are getting different results for the equilibrium solution and
the time evolution of the number of produced particles $c$.
 This is very transparent when
solving the rate equations (4) and (5).

In the limit when $\langle N_c \rangle \gg 1$, the standard
Eq.~(\ref{normal1}) is valid and has the well known  solution:
\ber \langle N_c \rangle ^{} (\tau)=\langle  N_c\rangle_{\rm
eq}^{\rm } \tanh \left ( \tau/\tau_0^{\rm } \right ), \eer  where
the equilibrium value $\langle N_c\rangle_{\rm eq}^{}$  of the
number of particles $c$
 and the relaxation time constant $\tau_0^{}$ are
given by: \ber \langle N_c\rangle_{\rm eq}^{\rm }= \sqrt
\epsilon~~, ~~ \tau_0^{\rm } = \frac {V}{L \sqrt \epsilon},
\label{eqgc} \eer respectively, with    $\epsilon\equiv G \langle
N_{a}\rangle\langle N_{a}\rangle/L$.

In the particular case when the particle momentum distribution is
thermal, the gain  ($G$) and loss  ($L$) terms  are just  the
thermal averages of the absorption and production cross sections
with \cite{our1}
 \ber
\frac{G}{L}= \frac {\langle N_c\rangle_{\rm eq}\langle N_{\bar
c}\rangle_{\rm eq}} {\langle N_a\rangle_{\rm eq}\langle N_{
b}\rangle_{\rm eq}}, \eer where we have employed the detailed
balance relation between   production $\sigma_{ab}$ and absorbtion
$\sigma_{c\bar c}$ cross section for $ab
 \to c\bar c$ processes.

Assuming a Boltzmann particle momentum distribution, the
equilibrium average number of particles $k$ is obtained after
momentum integration as:
 \ber \langle N_k\rangle_{\rm eq}^{\rm }= { {d_{k}}\over
{2\pi^2}} VTm_k^2  K_2 (m_k/T),
 \label{15}
\eer
 where $d_k$'s denote spin--isospin  degeneracy factors,
$ m_k$ the  particle mass, and $K_2$ is the modified Bessel
function.

This is a well known result for the average number of particles in
the    Grand Canonical (GC) ensemble with respect to U(1) charge
conservation. The chemical potential, which is usually present in
the GC ensemble, vanishes  in our case,  because of the
requirement of U(1) charge neutrality of a system. Thus, the
solution (\ref{normal1})
 results in the expected value for the equilibrium limit in GC
formalism where the  charge is conserved  on the average.

In the opposite limit, where $\langle N_c \rangle \ll 1$, the time
evolution of particle abundance is described by Eq.~(5), which has
the following solution: \ber \langle N_c \rangle ^{\rm C} (\tau)=
\langle N_c\rangle_{\rm eq}^{\rm C} \left ( 1-
e^{-\tau/\tau_0^{\rm C}} \right ),
 \label{16}
\eer with  the equilibrium value and relaxation time  given by
\ber \langle N_c\rangle_{\rm eq}^{\rm C}= \epsilon,~ \tau_0^{\rm
C} = \frac {V}{L}. \label{eqc} \eer

The above result is  the asymptotic limit of the canonical C
formulation of conservation laws \cite{our1}. Here the charge
related with U(1) symmetry is exactly and locally conserved,
contrary to the GC formulation where this conservation is only
valid on the average.

Comparing Eq.~(\ref{eqgc}) with Eq.~(\ref{eqc}), we first find
that, for  $\langle N_c \rangle \ll 1$, the equilibrium value in
the canonical formulation is far smaller than what is expected
from the grand canonical result  as

\ber \langle N_c\rangle_{\rm eq}^{\rm C}={\langle N_c\rangle_{\rm
eq}^{\rm }}^2 \ll {\langle N_c\rangle_{\rm eq}^{\rm }}.
 \eer

 Secondly, we
can  conclude  that the relaxation time for a canonical system is
far shorter than what is expected from the grand canonical value
as \ber \tau_0^{\rm C}=\tau_0^{\rm } \langle N_c\rangle_{\rm
eq}^{\rm } \ll \tau_0^{}, \eer
 since in the limit $\bf (ii)$  the equilibrium
value  $\langle N_c\rangle_{\rm eq}^{\rm } \ll 1$.

 The above discussion shows the
importance of the canonical description of quantum number
conservation if the U(1) charged particle multiplicity  is small.
We also see that the  volume dependence differs in the two cases.
The particle density in the GC limit is $V$-independent whereas,
in the opposite case, the density scales linearly with $V$.

 We note that $\bf (i)$ and
 $\bf (ii)$  limits are essentially determined by the size of
$\langle \delta N_c^2 \rangle$, the  fluctuation of the number of
particles  $c$. The grand canonical results correspond to small
fluctuations, i.e. $\langle \delta N_c^2 \rangle/\langle N_c
\rangle^2 \ll 1$, while the canonical description is relevant if
the
   fluctuations are large,  i.e.
$\langle \delta N_c^2 \rangle/\langle N_c \rangle^2 \gg 1$.

\section{Master equation - general solution}
In the previous section, we formulated a general master equation
(1) for the probability to find the number of U(1) charged
particles being produced by the  binary $ab\to c\bar c$ process.
The approximate solutions of Eq.~(1) were   considered in two
limiting cases. For a large number of produced particles carrying
U(1) charge, the time evolution and the equilibrium limit of
Eq.~(1) correspond to the GC result. We have also indicated the
difference between GC and asymptotic C results on the level of
rate equations for particle multiplicity. This difference is
particularly  transparent when comparing master equations for
probabilities.

\subsection{Master equation for abundant U(1) charged particle production}
For abundantly produced particles $c$ and $\bar c$ through  $ab\to
c\bar c$, process we do not need to worry about strong particle
correlations due to U(1) charge  conservation. This also means
that instead of imposing U(1) charge neutrality conditions through
$N_c-N_{\bar c}=0$,  one assumes conservation on the average, that
is $\langle N_c\rangle -\langle N_{\bar c}\rangle =0$. In this
case the master equation (1) can be simplified.

In the derivation of Eq.~(1) the absorption terms proportional to
$L$ were obtained by constraining the U(1) charge conservation to
be local and  exact. For the conservation on the average, the
transition probability from $N_c$ to the $(N_c-1)$ state is no
longer proportional to $(L/V) N_c^2$ but rather to $(L/V)
N_c\langle N_{\bar c}\rangle$, since the exact conservation
condition $N_c=N_{\bar c}$ is no longer  valid and the number of
$\bar c$ particles can only be    determined by its average value.
In the GC limit, the  master equation for the time evolution of
the probability $P_{N_c}(\tau )$ takes the following form:

\ber \frac{dP_{N_c}}{d\tau}&=&{G\over V} \langle
N_{a}\rangle\langle N_{b}\rangle P_{N_c-1} +\frac {L}{V}
(N_c+1)\langle N_{\bar c}\rangle P_{N_c+1}
 \nonumber \\
&-& {G\over V}\langle N_{a}\rangle \langle N_{b}\rangle
P_{N_c}~~~- \frac {L}{V} N_c\langle N_{\bar c}\rangle P_{N_c} .
\label{eq2} \eer

Multiplying the above equation by $N_c$, summing over $N_c$ and
using the condition that $\langle N_c\rangle =\langle N_{\bar
c}\rangle $, one recovers Eq.~(4), the rate equation for $\langle
N_c\rangle $ in the GC ensemble. The above equation is thus indeed
the general master equation for the probability function in the GC
limit. Comparing this equation with the more general Eq.~(1), one
can see that the main difference is contained in the absorption
terms, which are linear instead of quadratic in particle number.

 Equation (14) can be solved exactly. Indeed, introducing the
generating function $g(x,\tau)$ for $P_{N_c}$,

\be g(x, {\tau }) = \sum_{N_c=0}^{\infty} x^{N_c} P_{N_c}(\tau),
\ee the  iterative equation (14) for probability can be converted
into a differential equation for the generating function:
\begin{equation}
  \frac{\partial g(x,\tau)}{\partial \tau}={{L}\over V}\sqrt\epsilon (1-x)[g^\prime -{\sqrt\epsilon}
  g],
\end{equation}
with the general solution \cite{our2}:
\begin{equation}
  g(x,\tau)=g_0(1-xe^{-\tilde\tau})e^{\sqrt{\epsilon}(1-x)(e^{-\tilde\tau}-1)}, \label{solu}
  \end{equation}
where    $g' \equiv
\partial g / \partial x$,       $\tilde\tau=(L\sqrt\epsilon /V)\tau $ and
$\sqrt\epsilon =\langle N_c\rangle_{eq}$ given by Eq.~(9).

One can readily find the equilibrium solution to the above
equation. Taking the limit $\tau=\infty$ in Eq.~(17) leads to
\begin{equation}
 g_{\rm eq}(x)=e^{-\sqrt{\epsilon}(1-x)}, \label{eq}
\end{equation}
with the corresponding equilibrium multiplicity distribution:
\begin{equation}
  P_{N_c,{\rm
  eq}}=\frac{(\sqrt{\epsilon})^{N_c}}{{N_c}!}e^{-\sqrt{\epsilon}}.
\end{equation}
This  is a Poisson distribution with averaged multiplicity
$\sqrt{\epsilon}$.

\subsection{Equilibrium solution - general case}

The master equation (14),  describing the  evolution of the
probability function in the GC limit,  could be solved exactly.
The general equation (1), however, because  the quadratic
dependence of the absorption term, can only be solved
numerically. Nevertheless the equilibrium result for particle
multiplicity can be given.

Converting Eq.~(1)   for $P_{N_c}$'s into a partial differential
equation for the  generating function
\begin{equation}
g(x,\tau) = \sum_{N_c=0}^\infty x^{N_c} P_{N_c} (\tau). \label{23}
\end{equation}
one  finds \cite{our1}
 \ber \frac {\partial g(x,\tau)}{\partial
\tau} = \frac {L}{V} (1-x) \left (x g''+ g'- \epsilon g \right ).
\eer
The equilibrium solution  $g_{\rm eq}(x)$ thus obeys the following
equation: \ber xg_{\rm eq}'' + g_{\rm eq}' - \epsilon g_{\rm eq} =
0. \label{equil} \eer By a substitution of variables  ($x =
y^2/4/\epsilon$), this equation is   reduced to the Bessel
equation, with the following solution:

\ber g_{\rm eq}(x) = \frac {1}{I_0 ( 2\sqrt \epsilon)} I_0 (
2\sqrt {\epsilon x} ), \label{26} \eer where the normalization is
fixed by $g(1) = \sum P_{N_c} = 1$.

The equilibrium value for the probability function $P_{N_c}$ is
now expressed from      Eqs.~(20--23) as:

\ber P_{N_c,\rm eq}=\frac {\epsilon^{N_c}}{I_0 ( 2\sqrt \epsilon )
(N_c!)^2}. \eer

We note that the equilibrium distribution of the particle
multiplicity in not a Poissonian. This fact was first indicated in
equilibrium studies in \cite{peter}. In our case  this is a direct
consequence of the quadratic dependence on the multiplicity in the
loss terms of the master equation (1). A Poisson distribution is
obtained from Eq.~(24) if $\sqrt\epsilon \gg 1$, that is for large
particle multiplicity. In this case the C ensemble coincides with
the GC approximation.

The result for the  equilibrium average number of particles  $c $
 can be obtained  as:
  \ber \langle N_c \rangle_{\rm eq}= g'(1)= \sqrt
\epsilon \frac {I_1 ( 2\sqrt \epsilon)}{I_0 ( 2\sqrt \epsilon)}.
\label{neq} \eer

 The  equilibrium solution presented above coincides with the
expected result for the particle multiplicity in the canonical
ensemble with respect to U(1)  charge conservation
\cite{cleymans}. Thus, the rate equation formulated in Eq.~(1) is
valid for any arbitrary value of $\langle N_c\rangle $ {\it and}
obviously reproduces the standard grand canonical result for large
$\langle N_c\rangle$. We can study the chemical equilibration of
U(1) charged particles following Eq.~(1) independently from  the
thermal conditions in a system.

\subsection{Master equation for net U(1) charge in thermal environment}

So far, constructing the  evolution equation for probabilities, we
have assumed that  there is no  net U(1) charge  in a  system. In
the application of the statistical approach to particle production
in heavy ion collisions, the above assumption is only justified
when the initial state is U(1) charge neutral and when considering
particle yields in full phase space. However, because of
experimental limitations, one often deals with  data in restricted
kinematical windows. Here the overall U(1) charge is no longer
zero and a generalization of the above master equation is
required.

In the following we construct the evolution equation for
$P_{N_c}^S(t)$ in a thermal medium assuming that  its net U(1)
charge $S$ is non--vanishing. The presence of the net charge
requires a  modification of absorption terms in Eq.~(1). The
transition probability per unit time from the $N_c$ to the $N_c-1$
state was proportional to $(L/V)N_cN_{\bar c}$. Given an over all
net charge $S$ the exact  U(1) charge conservation implies that
$N_c-N_{\bar c}=S$. The transition probability from $N_c$ to
$N_c-1$ due to pair annihilation is thus $(L/V)N_c(N_{ c}-S)$.
Following the same procedure as in Eq.~(1) one can formulate the
following master equation for the probability $P_{N_c}^S(t)$ to
find $N_c$ particles $c$ in a thermal medium with a  net charge
$S$:

\ber \frac{dP_{N_c}^S}{d\tau}&=&{G\over V} \langle
N_{a}\rangle\langle N_{b}\rangle P_{N_c-1}^S +\frac {L}{V}
(N_c+1)(N_c+1-S) P_{N_c+1}^S
 \nonumber \\
&-& {G\over V}\langle N_{a}\rangle\langle  N_{b}\rangle
P_{N_c}^S~~~- \frac {L}{V} N_c(N_c-S) P_{N_c}^S , \label{eqS} \eer
which obviously reduces to Eq.~(\ref{eq1}) for $S=0$.

To get the equilibrium solution for the probability and
multiplicity, we again convert the above equation to the
differential form for the generating function $ g^S(x,\tau) =
\sum_{N_c=0}^\infty x^{N_c} P_{N_c}^S (\tau)$:

\ber \frac {\partial g^S(x,\tau)}{\partial \tau} = \frac {L}{V}
(1-x) \left (x g_S''+ g_S'(1- S)-\epsilon g_S \right ). \eer
In equilibrium, ${\partial g^S(x,\tau)}{\partial \tau}=0$ and the
solution for $g^S_{\rm eq}$ can be found as follows:

\ber g_{\rm eq}(x) = \frac {x^{S/2}}{I_S ( 2\sqrt \epsilon)} I_S (
2\sqrt {\epsilon x} ), \label{27} \eer
where   the normalization is fixed by $g(1) = \sum P_n = 1$.

The master equation for the probability to find $N_{\bar c}$
antiparticles $\bar c$, its corresponding differential form and
the equilibrium solution for the generating function can be
obtained by replacing $S\to -S$ in Eqs.~(26-28)



 The result for the  equilibrium average number of
particles  $\langle N_c\rangle_{\rm eq}$ and  antiparticles
$\langle N_{\bar c}\rangle_{\rm eq}$ is obtained from the
generating function using the relation:   $ \langle N_c
\rangle_{\rm eq}= g'(1)$. The final expressions read:

\ber \langle N_c \rangle_{\rm eq}=  \sqrt \epsilon \frac {I_{S-1}
( 2\sqrt \epsilon)}{I_S ( 2\sqrt \epsilon)}~~,~~ \langle N_{\bar
c} \rangle_{\rm eq}=  \sqrt \epsilon \frac {I_{S+1} ( 2\sqrt
\epsilon)}{I_S ( 2\sqrt \epsilon)}.
 \label{neq1} \eer
The strangeness conservation is explicitly seen by  taking the
difference of these equations, which yields, the net value of the
U(1) charge $S$.

The  results of Eq.~(\ref{neq1}) were  previously derived from the
equilibrium partition function by using the projection method
\cite{cleymans}. The master equation derived here allows a study
of the time evolution of particle production in a thermal medium
and the  approach towards chemical equilibrium. This equation
explicitly accounts for  U(1) charge conservation for both
strongly correlated and uncorrelated particles.

 The results presented here can be extended to the more general case
where there are different particle species carrying the U(1)
charge being produced in a thermal environment \cite{lin}.

\section{Concluding remarks}
 We have  discussed the kinetics and chemical equilibration of particles
carrying quantum numbers related with U(1) symmetry. Our analysis
was restricted to the particular case of one kind of particle and
antiparticle  being produced by the  binary process. However, this
example is of physics interest as it can be applied to the study
of such problems as the   equilibration of  heavy quarks  in QCD
plasma due to gluon--gluon fusion or strangeness and   baryon
production due to baryon--baryon and meson--meson interactions.
The last process is of relevance in the application of statistical
models to the description of particle production in low energy
heavy ion collisions \cite{our1,lin,cleymans,lin1}. We have
discussed the procedure that allows the construction of  the  rate
for time evolution of particle multiplicity and its probability
distribution. Our result is quite general and applicable, as well
as for abundantly and rarely produced U(1) charged particles. We
have indicated the important differences between  two limiting
thermodynamical situations. In the first case the system
approaches the equilibrium limit corresponding  to the grand
canonical approximation, in the second that corresponding to the
canonical exact value. Of particular interest are the master
equations for particle multiplicity distributions, which can be
used to study not only the time evolution of the particle number
but also all their higher moments \cite{our2}. This allows, for
instance, a
 study of the time evolution of U(1)
quantum number fluctuations \cite{lin,our2}, which has been
proposed as a possible signal for QCD phase transition
\cite{jeon}. \vskip 0.8cm

\centerline {\bf ACKNOWLEDGMENTS} \vskip 0.4cm
 Stimulating discussions with
M. Gyulassy, Z. Lin, H. Satz, M. Stephanov and Xin-Nian Wang are
kindly acknowledged. K.R. also acknowledges a partial support of
the Polish Committee for Scientific Research (KBN-2P03B 03018).
The work of V.K. was partially supported by GSI and partially by
the Director, Office of Science, Office of High Energy and Nuclear
Physics, Division of Nuclear Physics, and by the Office of Basic
Energy Sciences, Division of Nuclear Sciences, of the U.S.
Department of Energy under Contract No. DE-AC03-76SF00098.

\end{document}